\documentclass{article}
\usepackage[english]{babel}
\usepackage{threeparttable}

\usepackage[letterpaper,top=2cm,bottom=2cm,left=3cm,right=3cm,marginparwidth=1.75cm]{geometry}
\usepackage{amsmath}
\usepackage{graphicx}
\usepackage[colorlinks=true, allcolors=blue]{hyperref}
\usepackage{algorithm}
\usepackage[noend]{algpseudocode}
\usepackage{bm}	
\usepackage{indentfirst}
\usepackage{makecell}
\usepackage{caption}
\usepackage{graphicx}
\usepackage{float} 
\usepackage{subfigure}
\usepackage{pgfplots} 

\title{\textbf{DiffWA: Diffusion Models for Watermark Attack}}
\author{Xinyu Li\\Xi'an Jiaotong University,Xi'an,China\\2213311049@stu.xjtu.edu.cn}
\date{}

\begin{document}
\maketitle

\begin{abstract}
With the rapid development of deep neural networks(DNNs), many robust blind watermarking algorithms and frameworks have been proposed and achieved good results. At present, the watermark attack algorithm can not compete with the watermark addition algorithm. And many watermark attack algorithms only care about interfering with the normal extraction of the watermark, and the watermark attack will cause great visual loss to the image. To this end, we propose DiffWA, a conditional diffusion model with distance guidance for watermark attack, which can restore the image while removing the embedded watermark. The core of our method is training an image-to-image conditional diffusion model on unwatermarked images and guiding the conditional model using a distance guidance when sampling so that the model will generate unwatermarked images which is similar to original images. We conducted experiments on CIFAR-10 using our proposed models. The results shows that the model can remove the watermark with good effect and make the bit error rate of watermark extraction higher than 0.4. At the same time, the attacked image will maintain good visual effect with PSNR more than 31 and SSIM more than 0.97 compared with the original image.
\end{abstract}

\section{Introduction}
Blind watermarking is an invisible image watermark that can be used for copyright protection\cite{22,30}. With the development of DNNs, blind watermarking technology has made great progress. In 2018, Zhu et al.\cite{1} proposed an architecture for watermarking named HiDDeN, which was the first end-to-end watermark framework. Also in 2018, a framework for differential watermarking algorithm (ReDMark) was proposed by Ahmadi et al.\cite{29}. In 2020, Hao et al.\cite{2} completed the task of watermarking based on generative adversarial networks. Also in 2020, Lee et al.\cite{3} proposed a watermarking network without any resolution dependent layers or components to finish the task of watermarking. 

On the watermark attack side, researchers attack the watermark added to the image in various ways, attempting to make the watermark embedded in the image can not be extracted correctly. In 2018, a black box attack method based on adversarial learning for digital watermarking was proposed by Quiring et al.\cite{4}. In 2020, Nam et al.\cite{5} proposed a network named WAN (watermarking attack network) for watermark attack. By introducing residual dense blocks to the network, they allowed the proposed model to recognize both local and global features of images to remove the watermark. Geng et al.\cite{6} proposed a CNN-based real-time attack method against the robust watermarking algorithm, which is able to preprocess the image and destroy the watermark extraction without any prior knowledge. 

In recent years, researches on blind watermarking usually focus on watermarking addition, aiming to improve the robustness of proposed watermarking algorithm to protect copyright. Some watermarking algorithms can recover the watermark information with very low bit error rate or even lossless in the face of some existing watermarking attacks, which shows that the watermarking attack algorithm can no longer meet the requirements of the watermarking algorithm. Aiming to improve the performance of the watermarking model in the simulation attack, new watermark attack algorithms should be proposed.
  
Inspired by the success of image generative diffusion models, we proposed to introduce the diffusion models to the domain of watermark attack. Different from other generative models like Generative Adversarial Networks\cite{23}, diffusion models define an inference process to denoising the images from random noises. For Denoising Diffusion Probabilistic Models (DDPMs)\cite{7}, this process is based on a Markov chain while for Denoising Diffusion Implicit Models (DDIMs)\cite{8}, this process is non-Markov. In recent years, diffusion has been widely used in image editing\cite{24,25}, image inpainting\cite{26,27}, super resolution\cite{10,28}, and so on. It is natural to think of using the inference process of the diffusion models for watermark removal. Also, guided diffusion models and conditional diffusion models were proposed by Dhariwal \& Nichol\cite{9} and Saharia et al.\cite{10} to make the diffusion model generate images that meet certain requirements. As for watermark attack, the guided diffusion models and conditional diffusion models will allow the images after watermark attack to maintain high similarity with the original images. 

Therefore, we propose to use a conditional diffusion model with distance guidance, named DiffWA, to complete the task of watermark attack in this paper. We first trained the diffusion models using original images. Then at each step of inference process, the distance between generated images and watermarked images will be measured by a distance metric to guide the reconstructed images to be similar to the watermarked images and then to the original images. Meanwhile, because the generated images are generated under the condition of watermarked images and the model was trained on non-watermarked images, the reconstructed images will be closer to the original images and without watermark. Also, we propose one possible way to accelerate the inference process using an estimator and we try to combine two watermark attack models to get better watermark removal effect. In this paper, we adopted HiDDeN as the attacked watermarking scheme and we tested the results on CIFAR-10 dataset\cite{11}. The results shows that the proposed methods can remove the watermark at the bit error rate of extracted watermark about 0.4, and the highest 0.48. At the same time, the generated images have high similarity with the original images at the PSNR (peak signal to noise ratio) about 31 and SSIM (Structural Similarity)\cite{12} about 0.97.

\begin{figure}[htbp]
\centering
\includegraphics[width=0.9\textwidth]{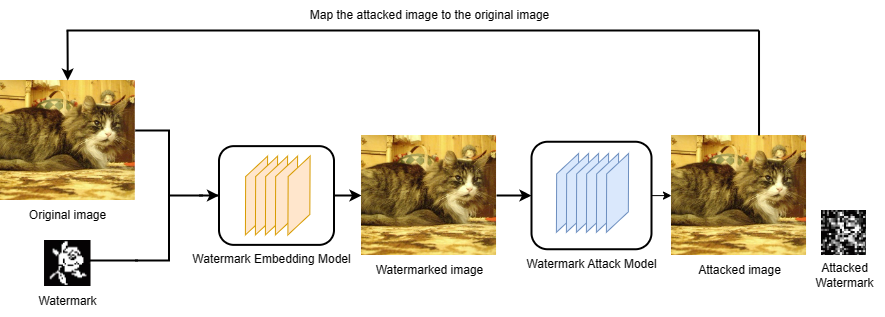}
\caption{Flowchart of the watermark attack models}
\end{figure}

\section{Background}

\subsection{HiDDeN}
Inspired by DNN's sensitivity to small perturbations in input images, Zhu et al.\cite{1} proposed the first end-to-end neural network for blind watermarking addition in 2018, named HiDDeN. HiDDeN consists of three parts, the encoder, the decoder and the discriminator. The inputs of the encoder are an original image and a string of message and it will output an encoded image. The decoder receives the encoded image and reconstructs the message encrypted into the encoded image. And the 
 aim of the discriminator is to determine if the image was encrypted by the encoder with the message, which plays the role of an adversary and will be cheated by encoder eventually. While training the network, the encoder and the decoder are trained jointly and the decoder will be fed both encoded images and distorted encoded images (encoded images after going through the noise layers) for training to make the watermark robust to various noises. It shows that the model has a fundamental advantage in robust watermarking with the encoded images able to resist a variety of watermarking attacks, like Gaussian blur, JPEG compression etc. 

\subsection{Denoising Diffusion Models}
Inspired by non-equilibrium thermodynamics\cite{13}, the denoising diffusion models was proposed. In the diffusion process of these models, random noise will be added to the images, which changes the real data distribution to a tractable Gaussian distribution. In the inference process, the model will reverse the diffusion process and learn how to remove the noise added to the images. Finally, the model will be able to generate images from randomly chosen noises and with some proper guidance, the model can generate some images that meet a specific need. 
\\
\\
\textbf{DDPM} In 2020, Ho et al.\cite{7} proposed the Diffusion Denoising Probabilistic Models (DDPMs). The diffusion process of DDPMs is a Markov process, which allows the noises to be added to the original image gradually. Let the $x_0 \sim p_{data}$, the latent variables${\ x}_{1,\ }{\ldots x}_T$ can be computed by following formula:

\begin{equation}
q(x_t| x_{t-1})=N(x_t;\sqrt{1-\beta_t}x_{t-1},\beta_tI)\label{1}
\end{equation}
where $\beta_t$ are predefined small positive constants. According to Ho et al., we define $\alpha_t=1-\beta_t,
{\bar{\alpha}}_t\ =\ \prod_{i=1}^{t}\alpha_t$, and we have $q\left(x_t\middle| x_0\right)=\ N\left(x_t,\ \sqrt{{\bar{\alpha}}_t}x_0,\left(1-{\bar{\alpha}}_t\right)I\right)$, Therefore, when T is large enough $x_t$ can be sampled by following equation:
\begin{equation}
x_t=\ \sqrt{{\bar{\alpha}}_t}x_0+\ \sqrt{1-{\bar{\alpha}}_t}\epsilon,\ where\ \epsilon\ a\ is\ a\ standard\ Gaussian\ noise\label{2}
\end{equation}
The inference process of DDPMs is also a Markov process. In this process, the model will estimate the noise and remove the noise added to the images. Let $x_T \sim N(0,I)$, the inference process from $x_T$ to $x_0$ can be defined as:
\begin{equation}
p_\theta\left(x_0,\ldots,x_{T-1}\middle| x_T\right)=\prod_{t=1}^{T}{p_\theta\left(x_{t-1}\middle| x_t\right),\ where\ p_\theta\left(x_{t-1}\middle| x_t\right)=N\left(x_{t-1};\mu_\theta\left(x_t,t\right),\sigma_t^2I\right)}\label{3}
\end{equation}
The mean $\mu_\theta\left(x_t,t\right)$ can be learned by a neural network and the variance $\sigma_t^2$ can be constants depended on timesteps\cite{7} or be learned by a neural network\cite{14}.
\\
\\
\textbf{DDIM} In 2021, Song et al.\cite{8} proposed Denoising Diffusion Implicit Models (DDIMs). Starting from $x_T\sim N\left(0,I\right)$ to clean image $x_0$, the inference process of DDIMs is a deterministic non-Markovian process, which can be defined as:
\begin{equation}
{\ x}_{t-1}=\sqrt{{\bar{\alpha}}_{t-1}}\left(\frac{x_t-\sqrt{1-{\bar{\alpha}}_t}\varepsilon_\theta\left(x_t,t\right)}{\sqrt{{\bar{\alpha}}_t}}\right)+\sqrt{1-{\bar{\alpha}}_{t-1}}\varepsilon_\theta\left(x_t,t\right)\label{4}
\end{equation}
where $\varepsilon_\theta\left(x_t,t\right)$ will be predicted by a neural network parameterized by $\theta$. 
\\
\\
\textbf{Guided Diffusion} Dhariwal \& Nichol\cite{9} introduced Adaptive Group Normalization (AdaGN) to diffusion models and used a classifier to guide the inference process of DDPMs and DDIMs to improve the quality and precision of sampling. To achieve this, the inference process of DDPM can be modified as: 
\begin{equation}
{x}_{t-1}=N(\mu+s\Sigma\nabla_{x_{t}}logp_{\phi}(y|x_{t}),\Sigma)\label{5}
\end{equation}
where $\mu$ and $\Sigma$ are the output of a diffusion model $\left(\mu_\theta\left(x_t,t\right),\Sigma_\theta\left(x_t,t\right)\right), p_\phi\left(y\middle| x_t\right)$ is the output of a classifier and y is the predicted label, s\ is the gradient scale. For DDIM, $\varepsilon_\theta\left(x_t,t\right)$ in Equation 4 will be replaced by$\ \hat{\varepsilon}$, which is defined as:
\begin{equation}
    \hat{\varepsilon}=\varepsilon_\theta\left(x_t,t\right)-\sqrt{1\ -\ {\bar{\alpha}}_t}\nabla_{x_t}logp_\phi\left(y\middle| x_t\right)\label{6}
\end{equation}
\\
\textbf{Conditional Diffusion} In 2021, an image-to-image diffusion framework named Palette was proposed by Saharia et al.\cite{15}. In Palette, the neural network will be fed in an image concatenated with a conditional image and output an image that meets certain requirements. In the inference process of this framework, $\varepsilon_\theta\left(x_t,t\right)$ will be replaced by $f_\theta\left(y,x_t,{\bar{\alpha}}_t\right)$, where $y$ is a conditional image.

\section{Proposed Methods}
\subsection{Preparations}

\noindent\textbf{Analysis of HiDDeN} The watermarking scheme attacked in this paper is HiDDeN\cite{1}. The region where the watermark is added to the original image determines the specific design of the watermark attack model. If the watermark is added in the high frequency domain of the image, it is necessary to reconstruct the high frequency domain of the image and retain the information of other domains as much as possible. Therefore, we first analyzed the region where the watermark is embedded. 
First, we performed Haar wavelet decomposition on the encoded image. Then, we set the corresponding frequency domain information after the wavelet decomposition to 0 respectively, reconstructed message and measured bit error rate (BER). Table 1 shows the results.

\begin{table}
\centering
\begin{tabular}{c c}
\hline
 The removed frequency composition & Bit Error Rate (BER) \\ \hline
 Low frequency component (LL) & 0.4761 \\  
 Horizontal high frequency component (LH) & 0.3633 \\
 Vertical high frequency component (HL) & 0.3850\\
 Diagonal high frequency component (HH) & 0.4023\\\hline
\end{tabular}
\caption{\label{tab:widgets}CIFAR-10 results for the BERs of the reconstructed image after removing a frequency component.}
\end{table}

Table 1 shows that HiDDeN relies on various frequency components to add watermark. The most dependent ones are the LL and HH parts, followed by HL and LH parts. Considering that every BER after removing one certain frequency component is relatively high, the watermark attack model should reconstruct both the low frequency and high frequency information while reconstructing watermarked images.
\\
\\
\textbf{Frequency View for Diffusion Models} In the diffusion process of diffusion models, Gaussian noise will be applied to the image at high and low frequencies. At the same time, the watermark will be destroyed. Also, it was proved by Yang et al.\cite{16} that in the inference process of diffusion, under the assumption of linearity, images are reconstructed from low frequency to high frequency. Therefore, with diffusion models trained on original images, we can recover the images from low frequency to high frequency using diffusion models, at the same time without watermark.

\subsection{DiffWA Framework}
The watermark attack model needs to restore the original image as much as possible while removing the watermark. Inspired by guided diffusion models and conditioanl diffusion models, we proposed to guide the conditional diffusion models with a distance metric in inference process, which allows the models generate images similar to original images or encoded images.
\\
\\
\textbf{Guided DDPM} Assumed that encoded image${\ x}_{en}=x+\delta$, where x is the original image, $\delta$ represents the watermark added to the original images. It was proved by Wang et al.\cite{17} and Nie et al.\cite{18} that in the inference process of DDPM, with the mean shifted by $-s\sum\mathrm{\nabla}_{x_t}D\left(x^t,x_{en}^t\right)$, where $\Sigma$ is the variance of$\ x_t$, $D$ is a distance metric, which can be MSE or –SSIM and s is a gradient scale, the generated picture $x$ could be guided to be similar with another picture $x_{en}$. Therefore, the inference process of DDPM can be modified as:
\begin{equation}
    x^{t-1}=N\left(\mu-s\sum\mathrm{\nabla}_{x^t}D\left(x^t,x_{en}^t\right),\Sigma\right),x_{en}^t=\sqrt{{\bar{\alpha}}_t}x_{en}+\sqrt{1-{\bar{\alpha}}_t}\epsilon,\epsilon\sim N(0,\mathbf{I})\label{7}
\end{equation}
where $\mu$ and $\mathrm{\Sigma}$ are the output of a diffusion model $\left(\mu_\theta\left(x_t,t\right),\mathrm{\Sigma}_\theta\left(x_t,t\right)\right)$
Also, the gradient scale is time-dependent, which is defined as:
\begin{equation}
    s_t=\frac{3\sqrt{1-{\bar{\alpha}}_t}}{\sqrt{{\bar{\alpha}}_t}\gamma}a\label{8}
\end{equation}
where $\gamma$ measures the bound of watermarking and $a$ is a chosen hyperparameter, which depends on the distance metric, the image resolution and the sampling method of diffusion models.
\\
\\
\textbf{Guided DDIM} The above derivation can be only applied to stochastic diffusion inference process and cannot be used for deterministic diffusion inference process, like DDIM\cite{8}. To end this, we adopted the score-based thick proposed by Song et al.\cite{19,20}. Assumed that we have a model $\varepsilon_\theta$, which is used for denoising, then it can be used in the score function:
\begin{equation}
    \nabla_{x^t}logp_\theta\left(x^t\right)=-\frac{1}{\sqrt{1-{\bar{\alpha}}_t}}\varepsilon_\theta\left(x^t\right)\label{9}
\end{equation}
In Equation 9, we can substitute $p_\theta$ for $p_{\theta,\phi}$,
\begin{equation}
\nabla_{x^t}logp_{\theta,\phi}\left(x^t\right)=\nabla_{x^t}logp_\theta\left(x^t\right)+\nabla_{x^t}logp_\phi\left(x_{en}^t\middle| x^t\right)\label{10}
\end{equation}
Here, we proposed a heuristic formula to approximate the probability 
\begin{equation}
    p_\phi\left(x_{en}^t\middle| x^t\right)=\frac{1}{Z}\left(1-tanh\left(D\left(x_{en}^t,x^t\right)\right)\right),\ where\ Z\ is\ a\ normalization\ factor\label{11}
\end{equation}
Finally, we can define ${\hat{\epsilon}}_\theta\left(x^t\right)$, which reflects the joint distribution:
\begin{equation}
    {\hat{\varepsilon}}_\theta=\varepsilon_\theta-\ \sqrt{1-{\bar{\alpha}}_t}s\nabla_{x^t}log\left(1-tanh\left(D\left(x_{en}^t,x^t\right)\right)\right)\label{12}
\end{equation}
where$\ D$ is a distance metric, s is a gradient scale, which is similar to s in Guided DDPM above.
Thus, we can replace the original $\varepsilon_\theta$ with ${\hat{\varepsilon}}_\theta$ to enable DDIM to conduct distance guidance in the inference process.
\\
\\
\textbf{Image-to-image Conditional Diffusion} The Palette\cite{15} framework of the form $p\left(x\middle| y\right)$ is trained to predict x under the conditional image y. Similarly, our watermark attack model is based on this framework, which could predict the original image x, under the conditional image $x_{en}$. A neural network $f_\theta$ is trained under the conditional image $x_{en}$ with the loss function:
\begin{equation}
    {E}_{\left(x,y\right)}{E}_{\epsilon \sim N\left(0,I\right)}{E}_{\bar{\alpha}}\Vert f_\theta(x_{en},\sqrt{\bar{\alpha}}\ x+\sqrt{1-\bar{\alpha}}\epsilon,\bar{\alpha})-\epsilon\Vert _p^p\label{13}
\end{equation}
In the inference process, with the conditional network $f_\theta$ instead of the unconditional network $\varepsilon_\theta$, the diffusion model can sample the images without watermarking under the condition of watermarking images, so as to ensure the high similarity of the output image compared with the original image. In conclusion, Algorithm 1 and Algorithm 2 summaries the proposed conditional diffusion sampling process with distance guidance using DDPM and DDIM.

\begin{figure}[t]
\centering
\includegraphics[width=0.8\textwidth]{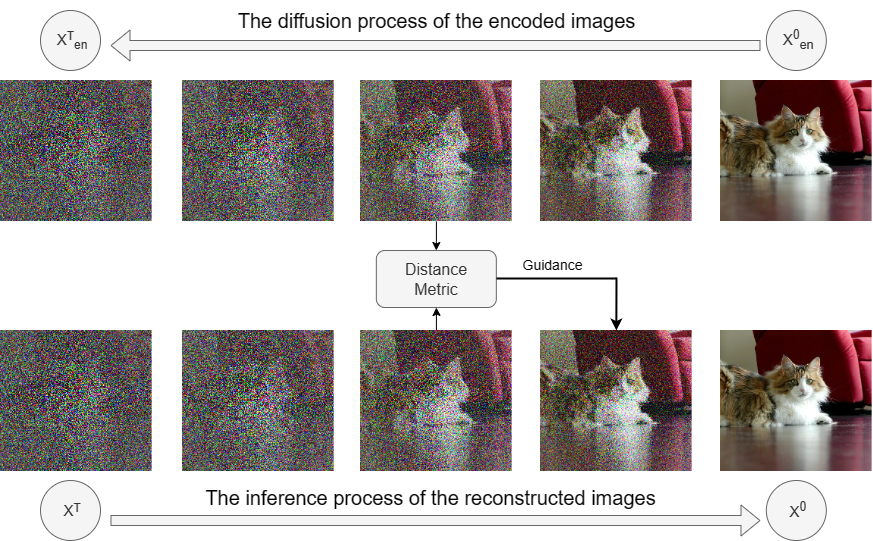}
\caption{How to guide the diffusion models with distance guidance}
\end{figure}

\begin{algorithm}
\renewcommand{\algorithmicrequire}{\textbf{Input:}}
\caption{Conditional diffusion sampling with distance guidance, given a DDPM $\left(\mu_\theta,\Sigma_\theta\right)$, gradient scale $s$, and conditional image $x_{en}$}

\textbf{Input: }Distance Metric Gradient, gradient scale $s$, encoded image $x_{en}$
\begin{algorithmic}[1]
\State{\textbf{for} $i \leftarrow 1$ \textbf{to} $M$ \textbf{do}}
\State{\indent The diffusion process: $x^t \leftarrow \sqrt{{\bar{\alpha}}_t}x_{en}+\sqrt{1-{\bar{\alpha}}_t}\epsilon$}
        \indent{\State{\indent{\textbf{for} $t \leftarrow T_c$ \textbf{to} $1$ \textbf{do}}}
        \State{\indent{\indent{$\mu,\Sigma\gets\mu_\theta\left(x_{en},x^t,{\bar{\alpha}}_t\right),\Sigma_\theta\left(x_{en},x^t,{\bar{\alpha}}_t\right)$}}}
        \State{\indent\indent$x^{t-1}\leftarrow sample\ from\ N\left(\mu-s\sum\mathrm{\nabla}_{x^t}D\left(x^t,x_{en}^t\right),\Sigma\right)$}
        \State{\indent\textbf{end for}}}
\State {\textbf{end for}}
\State {\textbf{return }$x^0$}
\end{algorithmic}
\end{algorithm}

\begin{algorithm}
\renewcommand{\algorithmicrequire}{\textbf{Input:}}
\caption{Conditional diffusion sampling with distance guidance, given a DDIM $\varepsilon_\theta$, gradient scale $s$, and conditional image $x_{en}$}

\textbf{Input: }Distance Metric Gradient, gradient scale $s$, encoded image $x_{en}$
\begin{algorithmic}[1]
\State{\textbf{for} $i \leftarrow 1$ \textbf{to} $M$ \textbf{do}}
\State{\indent The diffusion process: $x^t \leftarrow \sqrt{{\bar{\alpha}}_t}x_{en}+\sqrt{1-{\bar{\alpha}}_t}\epsilon$}
        \indent{\State{\indent{\textbf{for} $t \leftarrow T_c$ \textbf{to} $1$ \textbf{do}}}
        \State{\indent{\indent{${\hat{\varepsilon}}_\theta \leftarrow \varepsilon_\theta(x_{en},x^t,{\bar{\alpha}}_t)-\ \sqrt{1-{\bar{\alpha}}_t}s\nabla_{x^t}log\left(1-tanh\left(D\left(x_{en}^t,x^t\right)\right)\right)$}}}
        \State{\indent\indent${\ x}_{t-1} \leftarrow \sqrt{{\bar{\alpha}}_{t-1}}\left(\frac{x_t-\sqrt{1-{\bar{\alpha}}_t}{\hat{\varepsilon}}_\theta}{\sqrt{{\bar{\alpha}}_t}}\right)+\sqrt{1-{\bar{\alpha}}_{t-1}}{\hat{\varepsilon}}_\theta$}
        \State{\indent\textbf{end for}}}
\State {\textbf{end for}}
\State {\textbf{return }$x^0$}
\end{algorithmic}
\end{algorithm}

Here, we loop the denoising process for $M$ times to get better results for removing watermark and for each loop of denoising process, it is no need to sample from complete noise. We can distort the images as well as the watermark to step $T_c$ of the diffusion process, which makes the watermark ineffective, and then we only need to perform denoising in these $T_c$ steps to obtain the images without watermark. Also, $s$ can be set to 0 to let only conditional diffusion work and the conditional diffusion can be replaced by unconditional diffusion to let only distance guidance work.

\subsection{Estimator Acceleration}
 In order to accelerate the generation of the images without watermark, we introduced an estimator to this model. Assumed that $N$ is a time step in diffusion process, which is small compared to total steps $T$, $x^N$ is the original image after $N$ steps of diffusion. The estimator $f_e$ is used to fit the distribution of $x^N$, under the condition of $x_{en}$. Getting the output of estimator $x_e^N=f_e\left(x_{en}\right)$, we only need to perform $N$ steps of denoising on image $x_e^N$ to obtain the image without watermark. Simply, the estimator can be a ResNet\cite{21}. And Algorithm 3 shows the sampling using estimator.

\begin{figure}[htbp]
\centering
\includegraphics[width=0.8\textwidth]{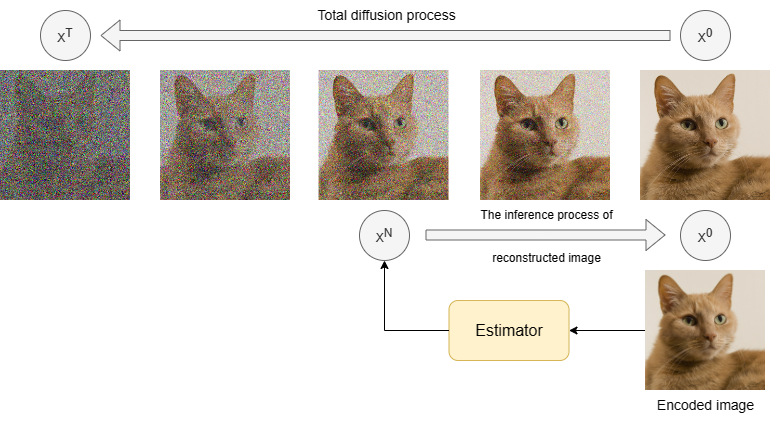}
\caption{How to accelerate the diffusion models using estimator}
\end{figure}
 
\begin{algorithm}
\renewcommand{\algorithmicrequire}{\textbf{Input:}}
\caption{Conditional Sampling using estimator, given an estimator $f_e$, a diffusion model, and encoded image $x_{en}$}

\textbf{Input: }Diffusion steps $N$, encoded image $x_{en}$
\begin{algorithmic}[1]
\State{$x_e^N \leftarrow f_e\left(x_{en}\right)$}
\State{\textbf{for} $t \leftarrow N$ \textbf{to} $1$ \textbf{do}}
\State{\indent{$x_e^{t-1}\ \leftarrow \ one\ denoising\ step\ for\ x_e^t$\\
\indent\indent\indent\indent\indent\indent$using\ conditional\ DDPM\ or\ DDIM\ with\ distance\ guidance$}}
\State {\textbf{end for}}
\State {\textbf{return }$x^0$}
\end{algorithmic}
\end{algorithm}

\subsection{Combinatorial Method}
In order to obtain better watermark removing effect, we can use a watermark attack model to preprocess the images, which shifts the encoded image distribution $x_{en}$ to a latent distribution $x_{latent}$. At this time, part of watermark is removed and the preprocessed images do not differ too much from the original images. Then we train the diffusion-based watermark attack model on $x_{latent}$ distribution. With this model, we can do further watermark removal and reconstruct the images more similar to original images. The preprocessing can be a proposed watermark attack framework or a diffusion-based watermark attack model.

\section{Experiments}
\subsection{HiDDeN}
To evaluate the proposed methods, we first trained a HiDDeN\cite{1} model on CIFAR-10\cite{11} training set with the message capacity metric BPP (Bits per pixel)=0.2. To increase the robustness of this watermarking algorithm, we combined the noisy layers available, including Crop layer ($p=0.035$), Cropout layer ($p=0.3$), Dropout layer ($p=0.3$), Gaussian blur layer, and JPEG compression. Aiming to enhance the ability for watermarking and scalability of this model, we introduced residual blocks\cite{21} to make the model wide enough and deep enough for watermarking on dataset. We trained the model for 40 epochs on training set on a RTX3060 until the image reconstruction loss less than 0.001, message reconstruction loss less than 0.001 on CIFAR-10 test set. 

\space\space We used PSNR (peak signal to noise ratio) and SSIM (structural similarity)[12] to measure the difference between the encoded images and the original images, which shows the function of the encoder, and used Bit accuracy to measure the ability of decoder to reconstruct the message. We tested the model with PSNR, SSIM on CIFAR-10 test set and Bit accuracy. Table 2 shows the results.
\begin{table}[b]
\centering
\begin{tabular}{c c}
\hline
Metric & Value \\\hline
PSNR & 32.62 \\
SSIM & 0.974 \\
Bit accuracy (original dataset) & 0.509 (random guess)\\
Bit accuracy (encoded dataset) & 0.999\\\hline
\end{tabular}
\caption{\label{tab:widgets}HiDDeN Evaluation results on CIFAR-10 test set}
\end{table}

We also measured the Bit accuracy under several distortions for encoded images to test the robustness of watermarking. PSNR and SSIM between the original images and distorted images. Table 3 shows the results.
\begin{table}
\centering
\begin{tabular}{c c c c c}
\hline
Distortion & PSNR &	SSIM &	Bit accuracy  \\\hline
Edge sharpening & 5.17 & 0.277 & 0.859 \\
Gaussian blur & 25.12 & 0.889 & 0.999  \\
Random noise $(U(0, 50))$ &	19.65 &	0.809 &	0.763 \\
Gaussian noise $(\sigma=20)$ & 21.62 &0.725 &0.740 \\
Salt and pepper noise $(p=0.1)$ & 15.10 &0.514&0.972\\
JPEG compression (quality=50)&26.08&0.903&	0.776 \\\hline
\end{tabular}
\caption{\label{tab:widgets}CIFAR-10 test results of encoded images under several distortions}
\end{table}

From Table 2 and Table 3, we knew that the HiDDeN model we trained has a good ability for watermarking and the watermark can resist several distortions, which lays the foundation for the following watermark attack experiments. It is worthing noting that the distortions used above are often be adopted as means of traditional watermark attack. The results also show that traditional watermark attack methods are difficult to work against HiDDeN.

\begin{figure}[htbp]
	\centering
        \subfigure{
        \includegraphics[width=3.5cm]{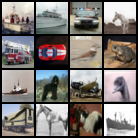}
        \includegraphics[width=3.5cm]{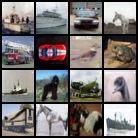}
        \includegraphics[width=3.5cm]{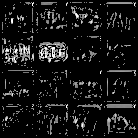}
        }
        \caption{HiDDeN-encoded results. From left to right are the original images, encoded images and 15 times the residual of the original and encode images}
 
\end{figure}

\subsection{Experiments on DiffWA}
Our conditional model is based on the image-to-image framework, Palette, proposed by Saharia et al.\cite{15}, which removes the class embedding of AdaGN\cite{9} layer. We set the total diffusion steps $T=1000$ in this paper. We trained the model with a batch size of 64  for forty thousand iterationson a RTX3060.

For comparison, an unconditional model was also trained using the architecture proposed by Dhariwal \& Nichol\cite{9}. The class embedding of AdaGN was also removed in this model. The training loss for this unconditional is same as $L_{simple}=E_{t,x,\epsilon}\Vert\varepsilon_\theta(\sqrt{\bar{\alpha}}\ x+\sqrt{1-\bar{\alpha}}\epsilon,t)-\epsilon\Vert_2^2$ proposed by Ho et al.\cite{7}, where $\varepsilon_\theta$ represents the diffusion model. The total diffusion steps and other training setting of the unconditional model were same as the conditional one.

In the inference process, for convenience, we defined $\eta=a/\gamma$ in equation 8. To get better performance, we set loop times $M=2$ for both DDPM and DDIM. For each loop, we set denoising steps $T_c=200$ for DDPM and $T_c=100$ for DDIM. In this experiment, we defined the distance metric as MSE (Mean Square Error) and -SSIM(Structural Similarity) respectively. Assume that the image has been normalized to a range of 0 to 1. When we used MSE as distance metric, for conditional models with distance guidance, we set $\eta=\ 0.05$ for DDPM and $\eta=-1$ for DDIM. For unconditional models with distance guidance, we set $\eta=6.25$ for DDPM and $\eta=-125$ for DDIM. When we used -SSIM as distance metric, for conditional models with distance guidance, we set $\eta=\ 255$ for DDPM and $\eta=-25500$ for DDIM. For unconditional models with distance guidance, we set $\eta=\ 63750$ for DDPM and $\eta=-6375000$ for DDIM. For models without distance guidance, $\eta$ was set to 0.

To measure the effect of watermark attack, we used SSIM and PSNR to measure the similarity of given images. We also evaluated the Bit accuracy between cleaned-images-reconstructed messages and original messages to measure watermark attack capability of the model. Table 4 shows the results of PSNR and SSIM. Table 5 shows the results of Bit accuracy.

Table 4 shows the model’s ability of reconstructing images and Table 5 shows the model’s ability of watermarking attack. From Table 5, conditional DDIM sampler with SSIM guidance performed best. And the image reconstruction capability of the model is reflected in its ability to make PSNR and SSIM between clean images and original images higher than PSNR and SSIM between encoded images and original images. In table 5, the conditional DDPM sampler without distance guidance showed the best ability of removing watermark. Also, we discovered that distance guidance can improve the similarity between clean images and original images while at the same time keep more watermark messages. Thus, the extent of distance guidance needs to be carefully determined to balance the similarity and the removal rate of watermark messages.

\begin{table}[htbp]
\centering
\begin{tabular}{c | c c c c c}
\hline
Method	&Sampler	&\thead{PSNR $\uparrow$\\ $({x}_{clean},x_{original})$}&\thead{	PSNR $\uparrow$ \\$({x}_{clean},x_{en})$}&\thead{	SSIM $\uparrow$ \\$({x}_{clean},x_{original})$}	&\thead{SSIM $\uparrow$ \\$({x}_{clean},x_{en})$}
 \\\hline
 & 	DDPM(MSE)&	27.89&	29.39&	0.924&	0.932\\
Only & DDIM(MSE)	&29.18&	30.94&	0.945&	0.947 \\\cline{2-6}
Guidance&DDPM(SSIM)&16.44&	16.53&	0.880&	0.891\\
 &DDIM(SSIM)&18.59&	18.77&	0.882&	0.898\\\hline
 
Only& DDPM&	31.06&	29.72&	0.963&	0.968\\
Conditional& DDIM&	28.05&	27.75&	0.944&	0.951\\\hline

 &DDPM(MSE)&	32.43&	31.84&	0.975&	0.980\\
Conditional&DDIM(MSE)&	32.42&	32.71&	0.980&	\textbf{0.982}\\\cline{2-6}
+Guidance&DDPM(SSIM)&32.44&	32.63&	0.975&	0.976\\
 &DDIM(SSIM)&	\textbf{33.09}&	\textbf{33.19}&	\textbf{0.981}&	0.980
\\\hline
\end{tabular}
\caption{\label{tab:widgets}CIFAR-10 test results of PSNR and SSIM. $x_{original}$represents the original images, $x_{en}$ represents the encoded images, $x_{clean}$ represents the images after watermarking attack, PSNR $(x_1,\ x_2)$ represents PSNR between image $x_1$ and $x_2$ and SSIM is represented in the same way.}
\end{table}

\begin{table}[H]
\begin{minipage}{0.5\linewidth}
\begin{tabular}{c | c c c }
\hline
Method	&Sampler	& Bit accuracy $\downarrow$	
 \\\hline
& 	DDPM(MSE)&	0.635\\
Only & DDIM(MSE)	&0.617\\\cline{2-3}
Guidance&DDPM(SSIM)&	0.789\\
&DDIM(SSIM)&	0.772\\\hline

Only& DDPM&	\textbf{0.549}\\
Conditional& DDIM&	0.587\\\hline

&DDPM(MSE)&	0.578\\
Conditional&DDIM(MSE)&	0.602\\\cline{2-3}
+Guidance&DDPM(SSIM)&	0.585\\
&DDIM(SSIM)&	0.584
\\\hline
\end{tabular}
\caption{\label{tab:widgets}CIFAR-10 test results of Bit accuracy.}
\end{minipage}
\hfill
\begin{minipage}{.5\linewidth}
\begin{tabular}{c | c c c }
\hline
Method	&Sampler	& Bit accuracy $\downarrow$	
 \\\hline

Only& DDPM&	0.592\\
Conditional& DDIM&	0.591\\\hline

&DDPM(MSE)&	0.584\\
Conditional&DDIM(MSE)&	0.595\\\cline{2-4}
+Guidance&DDPM(SSIM)&	0.564\\
&DDIM(SSIM)&	\textbf{0.549}&

\\\hline
\end{tabular}
\caption{\label{tab:widgets}Bit accuracy tested on CIFAR-10 test set using estimator acceleration}

\end{minipage}
\label{zrotate}
\end{table}

\pgfplotsset{compat=newest}
\begin{figure}[!h]
\centering
\subfigure[The curves of $\eta$ and SSIM, PSNR(left) and BER(right) using conditional DDPM with MSE guidance]{
\begin{tikzpicture}[scale=0.7]
\begin{axis}[
    axis y line*=left,
    xlabel=$\eta$, 
    ylabel=PSNR, 
    tick align=outside, 
    ]

\addplot[mark=*,blue] plot coordinates { 
(0,31.06)
(0.05,32.43)
(0.5,32.79)
(1,32.82)
(3,32.79)
(5,32.72)
(8,32.63)
(10,32.57)
};
\label{plot_one}

\end{axis}
\begin{axis}[
    axis y line*=right,
    axis x line=none,
    xlabel=$\eta$, 
    ylabel=SSIM,
    tick align=outside, 
    legend style={at={(0.84,0.2)},anchor=north} 
    ]
\addplot[mark=triangle,cyan] plot coordinates {
(0,0.963)
(0.05,0.975)
(0.5,0.976)
(1,0.977)
(3,0.977)
(5,0.976)
(8,0.976)
(10,0.975)
};
\addlegendimage{/pgfplots/refstyle=plot_one}\addlegendentry{PSNR}
\addlegendentry{SSIM}
\end{axis}
\end{tikzpicture}
\qquad
\begin{tikzpicture}[scale=0.7]
\begin{axis}[
    xlabel=$\eta$, 
    ylabel=BER,
    tick align=outside, 
    ]

\addplot[smooth,mark=*,blue] plot coordinates { 
(0,0.451)
(0.05,0.413)
(0.5,0.388)
(1,0.355)
(3,0.292)
(5,0.267)
(8,0.248)
(10,0.236)
};
\addlegendentry{BER}

\end{axis}
\end{tikzpicture}
}
\centering
\subfigure[The curves of $\eta$ and SSIM, PSNR(left) and BER(right) using conditional DDPM with SSIM guidance]{
\begin{tikzpicture}[scale=0.7]
\begin{axis}[
    axis y line*=left,
    xlabel=$\eta$, 
    ylabel=PSNR, 
    tick align=outside, 
    ]

\addplot[mark=*,blue] plot coordinates { 
(0,	31.06)
(100,	32.43)
(200,	32.44)
(255,	32.44)
(300,	32.44)
(500,	32.44)
(1000,	32.4)
(1500,	32.35)
};
\label{plot_one}

\end{axis}
\begin{axis}[
    axis y line*=right,
    axis x line=none,
    xlabel=$\eta$, 
    ylabel=SSIM, 
    tick align=outside, 
    legend style={at={(0.84,0.2)},anchor=north} 
    ]
\addplot[mark=triangle,cyan] plot coordinates {
(0,	0.963)
(100,	0.975)
(200,	0.9755)
(255,	0.9754)
(300,	0.9754)
(500,	0.9753)
(1000, 0.9747)
(1500, 0.974)

};
\addlegendimage{/pgfplots/refstyle=plot_one}\addlegendentry{PSNR}
\addlegendentry{SSIM}
\end{axis}
\end{tikzpicture}
\qquad
\begin{tikzpicture}[scale=0.7]
\begin{axis}[
    xlabel=$\eta$, 
    ylabel=BER, 
    tick align=outside, 
    ]

\addplot[smooth,mark=*,blue] plot coordinates { 
(0,	0.451)
(100,	0.431)
(200,	0.421)
(255,	0.415)
(300,	0.413)
(500,	0.4)
(1000, 0.367)
(1500, 0.339)

};
\addlegendentry{BER}

\end{axis}
\end{tikzpicture}
}

\centering
\subfigure[The curves of $\eta$ and SSIM, PSNR(left) and BER(right) using unconditional DDPM with MSE guidance]{
\begin{tikzpicture}[scale=0.7]
\begin{axis}[
    axis y line*=left,
    xlabel=$\eta$, 
    ylabel=PSNR, 
    tick align=outside, 
    ]

\addplot[mark=*,blue] plot coordinates { 
(1,	24.07)
(3,26.03)
(5,	27.37)
(6.25,	27.89)
(8,	28.17)
(10,	28.22)

};
\label{plot_one}

\end{axis}
\begin{axis}[
    axis y line*=right,
    axis x line=none,
    xlabel=$\eta$, 
    ylabel=SSIM, 
    tick align=outside, 
    legend style={at={(0.84,0.2)},anchor=north} 
    ]
\addplot[mark=triangle,cyan] plot coordinates {
(1,	0.83)
(3,0.897)
(5,	0.919)
(6.25,	0.924)
(8,	0.927)
(10,	0.931)

};
\addlegendimage{/pgfplots/refstyle=plot_one}\addlegendentry{PSNR}
\addlegendentry{SSIM}
\end{axis}
\end{tikzpicture}
\qquad
\begin{tikzpicture}[scale=0.7]
\begin{axis}[
    xlabel=$\eta$, 
    ylabel=BER, 
    tick align=outside, 
    ]

\addplot[smooth,mark=*,blue] plot coordinates { 
(1,	0.395)
(3,0.385)
(5,	0.373)
(6.25,	0.365)
(8,	0.351)
(10,	0.338)

};
\addlegendentry{BER}

\end{axis}
\end{tikzpicture}
}

\centering
\subfigure[The curves of $\eta$ and SSIM, PSNR(left) and BER(right) using unconditional DDPM with SSIM guidance]{
\begin{tikzpicture}[scale=0.7]
\begin{axis}[
    axis y line*=left,
    xlabel=$\eta$, 
    ylabel=PSNR, 
    tick align=outside, 
    ]

\addplot[mark=*,blue] plot coordinates { 
(10000,	14.89)
(30000,	16.14)
(50000,	16.42)
(63750,	16.44)
(80000,	16.46)
(100000,	16.49)
};
\label{plot_one}

\end{axis}
\begin{axis}[
    axis y line*=right,
    axis x line=none,
    xlabel=$\eta$, 
    ylabel=SSIM, 
    tick align=outside, 
    legend style={at={(0.84,0.2)},anchor=north} 
    ]
\addplot[mark=triangle,cyan] plot coordinates {
(10000,	0.7)
(30000,	0.828)
(50000,0.87)
(63750,	0.88)
(80000,	0.881)
(100000,	0.885)

};
\addlegendimage{/pgfplots/refstyle=plot_one}\addlegendentry{PSNR}
\addlegendentry{SSIM}
\end{axis}
\end{tikzpicture}
\qquad
\begin{tikzpicture}[scale=0.7]
\begin{axis}[
    xlabel=$\eta$, 
    ylabel=BER, 
    tick align=outside, 
    ]

\addplot[smooth,mark=*,blue] plot coordinates { 
(10000,	0.395)
(30000,0.29)
(50000,0.222)
(63750,	0.211)
(80000,	0.206)
(100000,	0.195)

};
\addlegendentry{BER}

\end{axis}
\end{tikzpicture}
}

\caption{The curves of $\eta$ and SSIM, PSNR and BER using (un)conditional DDPM with distance guidance}
\end{figure}
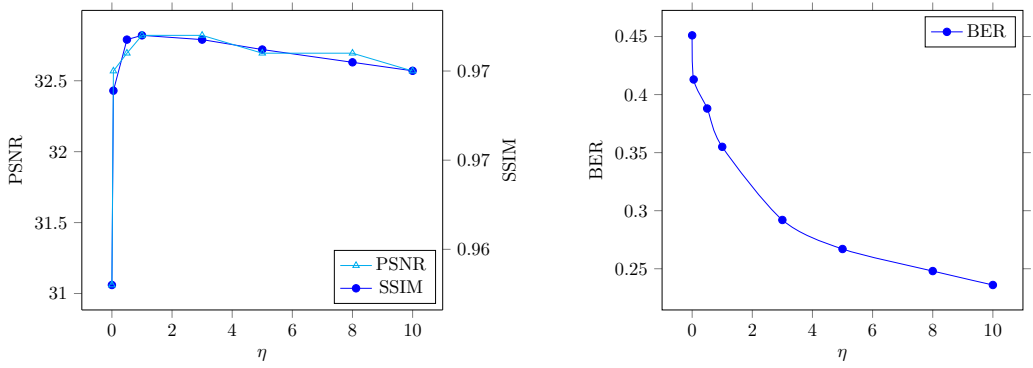
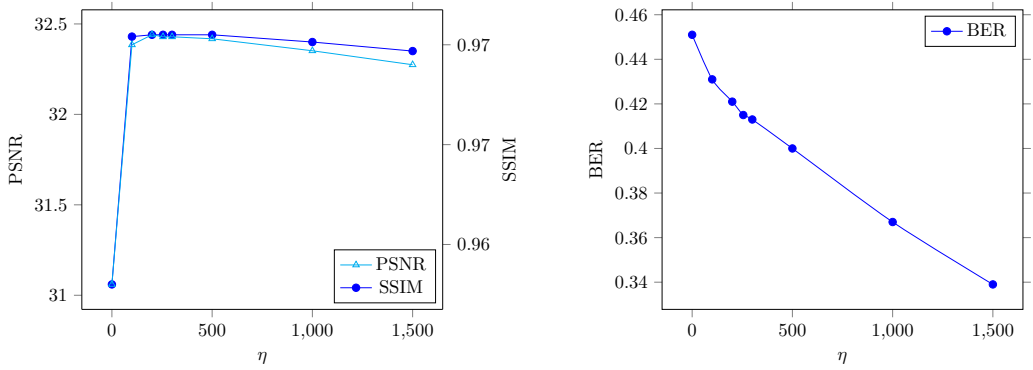
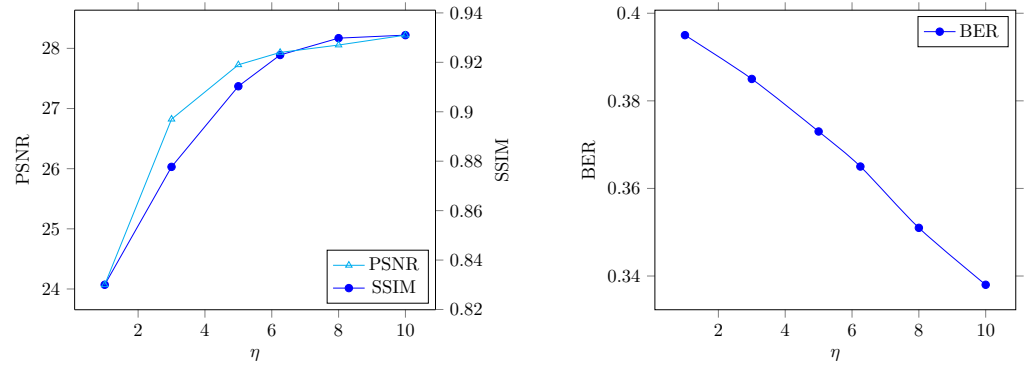
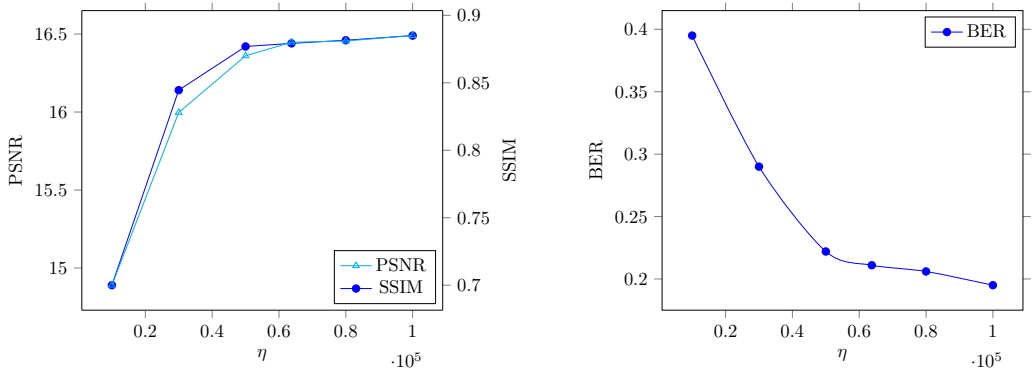

\begin{figure}[htbp]
	\centering

        \includegraphics[width=3.5cm]{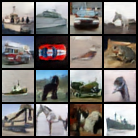}
        \includegraphics[width=3.5cm]{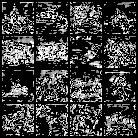}
        \includegraphics[width=3.5cm]{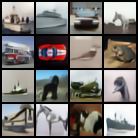}
        \includegraphics[width=3.5cm]{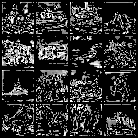}
        \subfigure[The results of Only distance guidance diffusion.The first line used MSE guidance and the second line used SSIM guidance.]{
        \includegraphics[width=3.5cm]{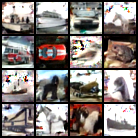}
        \includegraphics[width=3.5cm]{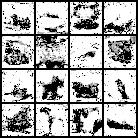}
        \includegraphics[width=3.5cm]{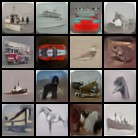}
        \includegraphics[width=3.5cm]{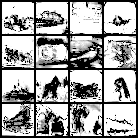}
        }
        \centering
        \subfigure[The results of Only conditional diffusion.]{
        \includegraphics[width=3.5cm]{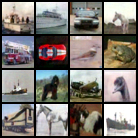}
        \includegraphics[width=3.5cm]{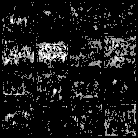}
        \includegraphics[width=3.5cm]{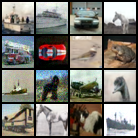}
        \includegraphics[width=3.5cm]{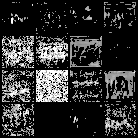}
        }
        \centering
        \includegraphics[width=3.5cm]{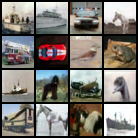}
        \includegraphics[width=3.5cm]{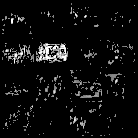}
        \includegraphics[width=3.5cm]{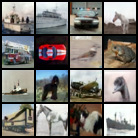}
        \includegraphics[width=3.5cm]{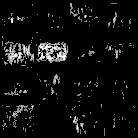}
        \subfigure[The results of conditional diffusion with distance guidance.The first line used MSE guidance and the second line used SSIM guidance.]{
        \includegraphics[width=3.5cm]{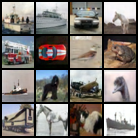}
        \includegraphics[width=3.5cm]{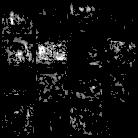}
        \includegraphics[width=3.5cm]{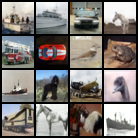}
        \includegraphics[width=3.5cm]{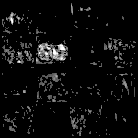}
        }
        \caption{The results of diffusion-based watermark attack. The first two columns are DDPM results of $x_{clean}$ and $15|x_{clean}-x_{original}|$. The last two columns are DDIM results of $x_{clean}$ and $15|x_{clean}-x_{original}|$. The original images are shown in Figure 1.}
 
\end{figure}

\noindent\textbf{The choice of $\eta$} Figure 5 shows the changes of PSNR and SSIM between clean images and original images and Bit Error Rate(BER) of messages extracted from clean images as $\eta$ increases. 

For conditional DDPM, when $\eta$ increases from zero to greater than 0, distance guidance starts to play a role in rapidly increasing PSNR and SSIM, which shows distance guidance helps to restore images. With the increase of $\eta$, PSNR and SSIM increase within a certain range and then they decrease slightly, which may be because the models rely too much on distance guidance and make the clean images and encoded images have similar PSNR and SSIM compared with the original images. Also, BER decreases with $\eta$ increases, which shows the side effect of distance guidance in reserving more messages and weakening the effect of watermark removal. Thus, the hyperparameter $\eta$ should be chosen carefully. Empirically, $\eta$ shouldn’t be too large, and it should be relatively close to 0.

For unconditional DDPM, with the increase of $\eta$, PSNR and SSIM increase monotonically and BER decreases monotonically. Comparing the conditional and unconditional models, the condition added to the models provide a basic function of restoring images and removing watermark. Also, the introduction of conditions reduces the sensitivity of the models to the choice of $\eta$ and enhances the robustness of the models.

\clearpage
\subsection{Estimator Acceleration}
We used ResNet34\cite{21} without pooling layers and fully connected layers as our estimator, which can map the encode images to original images after $N$ steps of diffusion. We set $N=100$ in this experiment, which means the diffusion model only needs 100 steps of denoising to get the final result. And we trained the estimator on a RTX3060 for 40 epochs on CIFAR-10 training set.

Here, to get better results, we just adopted the only conditional model and conditional model with distance guidance introduced in Experiments 4.2 for this experiment. Similarly, PSNR, SSIM, Bit accuracy was measured. Table 6 and Table 7 shows the results.

Table 6 and Table 7 shows that though using estimator acceleration, the model can still reconstruct the original images and remove the watermark with the performance similar to model without estimator acceleration.

\begin{figure}[htbp]
	\centering
        \subfigure[The results of Only conditional guidance diffusion using estimator.]{
        \includegraphics[width=3.5cm]{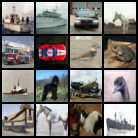}
        \includegraphics[width=3.5cm]{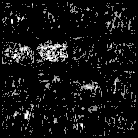}
        \includegraphics[width=3.5cm]{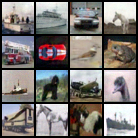}
        \includegraphics[width=3.5cm]{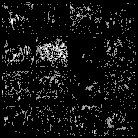}
        }
        \centering
        \includegraphics[width=3.5cm]{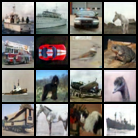}
        \includegraphics[width=3.5cm]{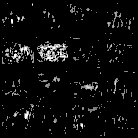}
        \includegraphics[width=3.5cm]{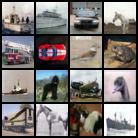}
        \includegraphics[width=3.5cm]{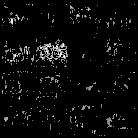}
        \subfigure[The results of conditional diffusion with distance guidance using estimator.The first line used MSE guidance and the second line used SSIM guidance.]{
        \includegraphics[width=3.5cm]{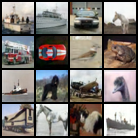}
        \includegraphics[width=3.5cm]{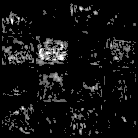}
        \includegraphics[width=3.5cm]{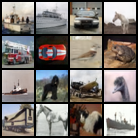}
        \includegraphics[width=3.5cm]{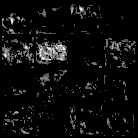}
        }
        \caption{The results of diffusion-based watermark attack using estimator. The first two columns are DDPM results of $x_{clean}$ and $15|x_{clean}-x_{original}|$. The last two columns are DDIM results of $x_{clean}$ and $15|x_{clean}-x_{original}|$. The original images are shown in Figure 1.}
 
\end{figure}

\begin{table}[htbp]
\centering
\begin{tabular}{c | c c c c c}
\hline
Method	&Sampler	&\thead{PSNR $\uparrow$\\ $({x}_{clean},x_{original})$}&\thead{	PSNR $\uparrow$ \\$({x}_{clean},x_{en})$}&\thead{	SSIM $\uparrow$ \\$({x}_{clean},x_{original})$}	&\thead{SSIM $\uparrow$ \\$({x}_{clean},x_{en})$}
 \\\hline

Only& DDPM&	31.57&	31.31&	0.966&	0.968\\
Conditional& DDIM&	31.87&	30.77&	0.963&	0.961\\\hline

&DDPM(MSE)&	32.61&	32.17&	0.978&	0.975\\
Conditional&DDIM(MSE)&	\textbf{33.30}&	31.79&	0.981& 0.976\\\cline{2-6}
+Guidance&DDPM(SSIM)&	32.44&	32.79&	0.975&	0.977\\
&DDIM(SSIM)&	33.01&	\textbf{32.98}&	\textbf{0.982}&	\textbf{0.980}

\\\hline
\end{tabular}
\caption{\label{tab:widgets}PSNR and SSIM tested on CIFAR-10 test set using estimator acceleration}
\end{table}

\subsection{Combinatorial Method}
  The output distribution of accelerated conditional diffusion model with MSE guidance using DDIM (the first model) in Experiments 4.3 was chosen to be the $x_{latent}$ for its better ability to restore the images. Then we trained another conditional diffusion model (the second model) under the condition of $x_{latent}$ using the training setting described in Experiments 4.2. PSNR, SSIM, Bit accuracy for this combinatorial method is shown in Table 8.

\begin{table}[H]
\begin{minipage}{0.5\linewidth}
   \centerline{\includegraphics[width=3.5cm]{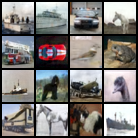}
        \includegraphics[width=3.5cm]{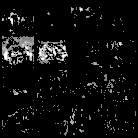}}
        {Figure 8: The results of diffusion-based watermark attack using combinatorial method. Left is $x_{clean}$ and right is $15|x_{clean}-x_{original}|$.The original images are shown in Figure 1.}
\end{minipage}
\hfill
\begin{minipage}{.4\linewidth}
\begin{tabular}{c c}
\hline
Metric & Value \\\hline
PSNR (${x}_{clean},x_{original}$)&	31.85 \\
PSNR (${x}_{clean},x_{en}$)&	29.17 \\
SSIM (${x}_{clean},x_{original}$)&	0.978\\
SSIM (${x}_{clean},x_{en}$)&	0.961\\
Bit accuracy&	0.514\\
\hline
\end{tabular}
\caption{\label{tab:widgets}CIFAR-10 test results for combinatorial method. The second model samples using conditional DDIM with MSE guidance.}
\end{minipage}
\label{zrotate}
\end{table}

The combinatorial method gives the best ability for watermark removal in experiments. Also, it can reconstruct the images with relatively high quality. 

\section{Conclusion}
We propose to use a conditional diffusion model with distance guidance for watermark attack in this paper, which shows good ability for watermark removal and image restoration. At the same time, with an estimator, we propose a possible way to speed up the inference process of these watermark attack models. Also, a combinatorial method is proposed to get better watermark removal effect. Future work may focus on how to restore images with a higher degree of fidelity and how to accelerate the proposed methods. In addition, we need more studies to prevent these proposed methods from misusing for copyright infringement. And it is necessary to analyze which watermarking techniques can resist this attack.

\bibliographystyle{acm}
\bibliography{reference}

\begin{thebibliography}{10}

\bibitem{29}
{\sc Ahmadi, M., Norouzi, A., Karimi, N., Samavi, S., and ReDMark, E.~A.}
\newblock Framework for residual diffusion watermarking based on deep networks.
\newblock {\em Expert Systems with Applications 146\/} (2020), 113157.

\bibitem{25}
{\sc Couairon, G., Verbeek, J., Schwenk, H., and Cord, M.}
\newblock Diffedit diffusion-based semantic image editing with mask guidance
  [preprint]. arxiv.
\newblock preprint, 2022.

\bibitem{9}
{\sc Dhariwal, P., and Nichol, A.}
\newblock Diffusion models beat gans on image synthesis.
\newblock {\em Advances in Neural Information Processing Systems 34\/} (2021),
  8780--8794.

\bibitem{6}
{\sc Geng, L.~F., Zhang, W.~M., Chen, H.~Z., Fang, H., and NH., Y.}
\newblock Real-time attacks on robust watermarking tools in the wild by cnn.
\newblock {\em Journal of Realtime Image Processing 17}, 3 (2020), 631--641.

\bibitem{23}
{\sc Goodfellow, I., Pouget-Abadie, J., Mirza, M., Xu, B., DavidWarde-Farley,
  S.~O., Courville, A., and Bengio, Y.}
\newblock Generative adversarial nets.
\newblock {\em In Advances in neural information processing systems\/} (2014),
  2672--2680.

\bibitem{2}
{\sc Hao, K.~L., Feng, G.~R., and XP., Z.}
\newblock Robust image watermarking based on generative adversarial network.
\newblock {\em China Communications 17}, 11 (2020), 131--140.

\bibitem{21}
{\sc He, K., Zhang, X., Ren, S., and Sun, J.}
\newblock Deep residual learning for image recognition.
\newblock In {\em Proceedings of the IEEE conference on computer vision and
  pattern recognition\/} (2016), pp.~770--778.

\bibitem{7}
{\sc Ho, J., Jain, A., and Abbeel, P.}
\newblock Denoising diffusion probabilistic models.
\newblock {\em Advances in Neural Information Processing Systems 33\/} (2020),
  6840--6851.

\bibitem{11}
{\sc Krizhevsky, A., Nair, V., and Hinton, G.}
\newblock Cifar-10 (canadian institute for advanced research).

\bibitem{3}
{\sc Lee, J.~E., Seo, Y.~H., and DW., K.}
\newblock Convolutional neural network-based digital image watermarking
  adaptive to the resolution of image and watermark.
\newblock {\em Applied Sciences 10\/} (2020), 19.

\bibitem{26}
{\sc Lugmayr, A., Danelljan, M., Romero, A., Yu, F., Timofte, R., and Gool,
  L.~V.}
\newblock Repaint inpainting using denoising diffusion probabilistic models.
\newblock In {\em Proceedings of the IEEECVF Conference on Computer Vision and
  Pattern Recognition\/} (2022), pp.~11461--11471.

\bibitem{24}
{\sc Luo, S., and Hu., W.}
\newblock Diffusion probabilistic models for 3d point cloud generation.
\newblock In {\em Proceedings of the IEEECVF Conference on Computer Vision and
  Pattern Recognition\/} (2021), pp.~2837--2845.

\bibitem{22}
{\sc Ma, Z.~H., Zhang, W.~M., Fang, H., Dong, X.~Y., Geng, L.~F., and NH., Y.}
\newblock Local geometric distortions resilient watermarking scheme based on
  symmetry.
\newblock {\em IEEE Trans. on Circuits and Systems for Video Technology 31}, 12
  (2021), 4826--4839.

\bibitem{5}
{\sc Nam, S.~H., Yu, I.~J., Mun, S.~M., Kim, D., and W., A.}
\newblock Wan watermarking attack network.
\newblock 2020.

\bibitem{14}
{\sc Nichol, A., and Dhariwal, P.}
\newblock Improved denoising diffusion probabilistic models. arxiv.
\newblock preprint, 2021.

\bibitem{18}
{\sc Nie, W., Guo, B., Huang, Y., Xiao, C., Vahdat, A., and Anandkumar, A.}
\newblock Diffusion models for adversarial purification. arxiv.
\newblock preprint, 2022.

\bibitem{4}
{\sc Quiring, E., and Rieck, K.}
\newblock Adversarial machine learning against digital watermarking.
\newblock In {\em Proc. of the 26th European Signal Processing Conf.
  (EUSIPCO)\/} (Rome, 2018), IEEE, pp.~519--523.

\bibitem{27}
{\sc Romero, A., Castillo, A., Abril-Nova, J., Timofte, R., Das, R., Hira, S.,
  Pan, Z., Zhang, M., Li, B., He, D., et~al.}
\newblock Ntire 2022 image inpainting challenge report.
\newblock In {\em Proceedings of the IEEECVF Conference on Computer Vision and
  Pattern Recognition\/} (2022), pp.~1150--1182.

\bibitem{15}
{\sc Saharia, C., Chan, W., Chang, H., Lee, C.~A., Ho, J., Salimans, T., Fleet,
  D.~J., and Norouzi, M.}
\newblock {\em Palette Image-to-image diffusion models}.
\newblock In NeurIPS 2021 Workshop on Deep Generative Models and Downstream
  Applications, 2021.

\bibitem{10}
{\sc Saharia, C., Ho, J., Chan, W., Salimans, T., Fleet, D.~J., and Norouzi,
  M.}
\newblock Image super-resolution via iterative refinement.
\newblock preprint, 2021.

\bibitem{13}
{\sc Sohl-Dickstein, J., Weiss, E., Maheswaranathan, N., and Ganguli, S.}
\newblock Deep unsupervised learning using nonequilibrium thermodynamics.
\newblock In {\em International Conference on Machine Learning\/} (2015),
  pp.~2256--2265.

\bibitem{8}
{\sc Song, J., Meng, C., and Ermon, S.}
\newblock Denoising diffusion implicit models.
\newblock In {\em International Conference on Learning Representations\/}
  (2021).

\bibitem{19}
{\sc Song, Y., and Ermon, S.}
\newblock Generative modeling by estimating gradients of the data distribution,
  2020.

\bibitem{20}
{\sc Song, Y., Sohl-Dickstein, J., Kingma, D.~P., Kumar, A., Ermon, S., and
  Poole, B.}
\newblock {\em Score-based generative modeling through stochastic differential
  equations}.
\newblock 2020.

\bibitem{28}
{\sc Wang, C., Yeo, K., Jin, X., Codas, A., Klein, L.~J., and Elmegreen, B.}
\newblock S3rp self-supervised super-resolution and prediction for
  advection-diffusion process. arxiv.
\newblock preprint, 2021.

\bibitem{17}
{\sc Wang, J., Zhang, Y., Li, Y., Liu, Y., and Zhang, X.}
\newblock Guided diffusion model for adversarial purification. arxiv.
\newblock preprint, 2022.

\bibitem{12}
{\sc Wang, Z., Bovik, A.~C., Sheikh, H.~R., and Simoncelli, E.~P.}
\newblock Image quality assessment from error visibility to structural
  similarity. ieee trans.
\newblock {\em Image Process 13}, 4 (2004), 600--612.

\bibitem{16}
{\sc Yang, Y., and Zhou, Y.}
\newblock Diffusion probabilistic model made slim. arxiv.
\newblock preprint, 2022.

\bibitem{30}
{\sc Yue, Z., Li, Z.~C., Yang, Y.~X., You, F.~C., and Liu~FP., A.}
\newblock histogram-based 2bin m-ary image digital watermarking algorithm.
\newblock {\em Acta Electronica Sinica 48}, 3 (2020), 531--537.

\bibitem{1}
{\sc Zhu, J.~R., Kaplan, R., Johnson, J., and HiDDeN, F.-F.~L.}
\newblock Hiding data with deep networks.
\newblock In {\em Proc. of the 15th European Conf. on Computer Vision (ECCV)\/}
  (Munich, 2018), Springer, pp.~682--697.

\end{thebibliography}
 
\end{document}